# Electronic Structure of Noncentrosymmetric Superconductor Li$_2$(Pd$_x$Pt$_{1-x}$)$_3$B Studied by Photoemission Spectroscopy

Rikiya YOSHIDA[1], Hiroyuki OKAZAKI[1], Mitsutoshi TAJIMA[1], Takayuki MURO[2], Izumi HASE[3], Kozo OKADA[1], Hiroyuki TAKEYA,[4] Kazuto HIRATA[4] Masaaki HIRAI,[1] Yuji MURAOKA,[1, 5] and Takayoshi YOKOYA[1, 5]

[1]*The Graduate School of Natural Science and Technology, Okayama University, 3-1-1 Tsushimanaka, Okayama 700-8530, Japan*
[2]*Japan Synchrotron Radiation Research Institute (JASRI)/SPring-8, 1-1-1 Kouto, Sayo, Hyogo 679-5198, Japan*
[3]*Nanoelectronics Research Institute, National Institute of Advanced Industrial Science and Technology, 1-1-4 Umezono, Tsukuba, Ibaraki 305-8568, Japan*
[4]*National Institute for Material Science, 1-2-1 Sengen, Tsukuba, Ibaraki 305-0047, Japan*
[5]*JST, CREST, 3-1-1 Tsushima-naka, Okayama 700-8530, Japan*

We have performed x-ray photoemission spectroscopy on the system of noncentrosymmetric superconductor, Li$_2$(Pd$_x$Pt$_{1-x}$)$_3$B. For Li$_2$Pt$_3$B, we found 2 major peaks with 2 other weak components, and the band calculations were in agreement with the observation. The assignment of valence band features using the calculated partial density of states determined that Pt 5$d$ and B 2$p$ contribute to the density of states at the Fermi level. The effect of antisymmetric spin-orbit coupling on the band structure might have been probed, and the analysis on the effect of Pt incorporation into the system indicates the smooth evolution of electronic structures. We presented the measurements of core levels (Pd 3$d$, Pt 4$f$, and B 1$s$) and discussed the chemical bonding states and electronic structures from them.

## 1. Introduction

Recently, superconductors without inversion symmetry, certainly a new topic on superconductivity, have attracted significant attentions. For the superconductors studied in the past, such as conventional BCS type superconductor, high-$T_c$ cuprates, or heavy fermion superconductors, symmetry of Cooper-pair has classified into either spin-singlet state or spin-triplet state. However, theoretical studies suggest that broken inversion symmetry in the crystal structure causes finite antisymmetric spin-orbit coupling[1,2] and allows the coexistence of spin-singlet and spin-triplet states in noncentrosymmetric superconductors.[3] Due to this anomalous behavior of noncentrosymmetric superconductors, the discovery of first noncentrosymmetric heavy fermion compound CePt$_3$Si (Ref.4) has stimulated both theoretical and experimental studies, having suggested unconventional superconductivity in it.[5-7] Currently, other noncentrosymmetric superconductors such as UIr,[8] CeIrSi$_3$,[9] CeRhSi$_3$,[10] and Li$_2$(Pd$_x$Pt$_{1-x}$)$_3$B [11,12] are known, but the role of antisymmetric spin-orbit coupling coupled to the absence of inversion symmetry must be investigated further to understand the symmetry of Cooper-pairs in noncentrosymmetric superconductors.

The system of transition metal compound Li$_2$(Pd$_x$Pt$_{1-x}$)$_3$B (space group: $P4_332$)[13] has attracted much attentions because two end compounds, Li$_2$Pd$_3$B and Li$_2$Pt$_3$B, exhibit completely different superconducting properties.[14-17] While experimental studies suggest that Li$_2$Pd$_3$B is in spin-singlet state[14] and Li$_2$Pt$_3$B is in spin-triplet state,[15] the superconducting critical temperature ($T_c$) linearly decreases as more platinum is incorporated into the system.[12] One could propose several scenarios to describe the superconductivity in Li$_2$(Pd$_x$Pt$_{1-x}$)$_3$B. For example, superconductivity related to the strong electron correlation effects concerned for the case of Sr$_2$RuO$_4$ (Ref. 18), or UPt$_3$ (Ref. 19) must be attended. In the previous studies, NMR,[14,15] specific heat measurements[20] and Photoemission spectroscopy (PES)[21] have implied that there is no indication of magnetic orders or electron correlations in Li$_2$Pd$_3$B. Similar effect is expected in Li$_2$Pt$_3$B, but some encourage direct observation of its electronic structure for rigorous evaluation as no PES experiment was done on Li$_2$Pt$_3$B. Others would discuss platinum's stronger antisymmetric spin-orbit coupling (ASOC) coupled with the absence of inversion symmetry for the cause of spin-triplet state in Li$_2$Pt$_3$B as mentioned in several literatures,[17,22] but this also requires further experimental evidences. The important point is that yet-to-be-defined superconductivity in the system of Li$_2$(Pd$_x$Pt$_{1-x}$)$_3$B has invoked several questions, and the settlement of them can deepen the understanding of superconductivity without inversion symmetry.

In this report, we have tried to approach the information on electronic structure in Li$_2$(Pd$_x$Pt$_{1-x}$)$_3$B through photoemission spectroscopy since the information can help the understanding of interesting superconductivity in it. Mainly, this work is committed to investigate the valence band of Li$_2$Pt$_3$B and to examine the effect of platinum substitution through the measurements of Li$_2$Pd$_{1.5}$Pt$_{1.5}$B. We found that the valence band of Li$_2$Pt$_3$B is in agreement with first principle calculations, and the observed valence band features were assigned using the calculated partial density of states (DOS), including the discussion near the Fermi level ($E_F$). We also found that the introduction of platinum into the system of $Li_2(Pd_xPt_{1-x})_3B$ causes smooth evolution of electronic structures, showing the smooth evolution of bandwidth. Core level (Pd 3*d*, Pt 4*f*, and B 1*s*) analysis of $Li_2Pd_{1.5}Pt_{1.5}B$ and $Li_2Pt_3B$ is discussed, and a Hubbard model calculation gives upper bound estimate of *U/W* ratio (*U*: on-site Coulomb potential, *W*: bandwidth) in $Li_2Pt_3B$. At the end of this paper, we give a brief comment on the anomalous superconductivity in $Li_2(Pd_xPt_{1-x})_3B$ based on our results.

## 2. Experimental

We prepared polycrystalline samples of $Li_2(Pd_xPt_{1-x})_3B$ for x = 0, 0.5, and 1 by an arc-melting method, and its detail is described elsewhere.[11, 12] Samples were confirmed to be in single phase by XRD measurements, and bulk superconductivity ($T_c$ = 7.1, 4.1, 2.8 for x = 0, 0.5, 1, respectively) was also checked by Meissner signals, showing sharp transitions. In addition, we measured a polycrystalline platinum plate (99.98%, The Nilaco Corporation) for comparison.

Photoemission measurements were performed at BL25SU and BL27SU of SPring-8 employing a hemispherical electron analyzer. We employed monochromatized synchrotron radiation of 1300 eV for photoelectron excitations, and measurements were taken at 13 K under $5.5 \times 10^{-8}$ Pa with a total energy resolution of 260 meV. For $Li_2Pd_{1.5}Pt_{1.5}B$ and $Li_2Pt_3B$, their clean surfaces were prepared by *in situ* fracture, as was checked by the absence of carbon 1*s* and oxygen 1*s* signals. Spectra for $Li_2Pd_3B$ were consistent with the ones in Ref 21. The platinum plate was scraped for surface preparation. The measured spectrum did not show any observable changes over the duration of experiments. We measured $E_F$ and the Au $4f_{7/2}$ peak of a thin gold film deposited near the samples to calibrate the binding energies of the sample spectra.

We followed the formulation in Ref. 23 and calculated a valence band spectrum of $Li_2Pt_3B$ using the full-potential augmented plane wave method (FLAPW) with the local density approximation. For this calculation, the computer program KANSAI-94 and TSPACE were utilized on the scalar relativistic scheme.[24] Since lattice constants and space group have been found experimentally,[13] we employed them in this calculation. For the value of the muffin-tin radius of each atom, we used 2.2 Bohr for lithium, 2.0 Bohr for platinum and 1.4 Bohr for boron. We imposed the wave vectors of the basis functions to satisfy $| k + G | < K_{max} = 6.90a$ (*k*: wave vector in the Brillouin zone, *G*: reciprocal lattice vector), outputting about 1360 augmented plane waves. In order to compare this calculation with experimental data, calculated partial DOS were summed taking photoionization cross sections[25] into account. Then, the sum was convoluted with Lorenzian function whose full-width-half maximum (FWHM) is given by $\alpha |E - E_F|$ and then with Gaussian function giving FWMH of 260 meV. Since Ref. 21 reported $\alpha = 0.13$ reproduced the measured spectra well, we employed the same value in this study.

## 3. Results and Discussion

Figure 1 illustrates the valence band photoemission spectra of samples measured at 13K with 1300eV photon energy, and spectra are normalized to the area. In the valence band of $Li_2Pt_3B$, we observed the following features: a sharp Fermi edge (features A), lower intensity at $E_F$ being than that of platinum, two major peaks at about 2.1eV and 4.7eV (features B and C), a weak component at around 6.5 eV (feature D), and another weak component (feature E) at 9.5eV. We compared these features to the result of FLAPW calculation on $Li_2Pt_3B$ shown in the bottom of Fig. 1 with a dotted line. The features A, B, C, D, and E in the observed spectrum are reproduced by the first principle calculation, and therefore, we suggest that feature E is not a correlation induced satellite, which was discussed elsewhere in the case of transition metal oxide.[26] Further comparison can be made to Fig. 2, which describes partial DOS. We find that feature B and C predominantly originate in Pt $5d$, and D is the hybridized states of Pt $5d$ and B $2p$. The feature E has B $2s$ and Pt $5d$ characters. Analogous to the case of $Li_2Pd_3B$,[22] we found that major contribution to the DOS at $E_F$ is from Pt $5d$ and B $2p$. The DOS at $E_F$ is 136.61 states/Ry per unit cell, being slightly higher than the value for $Li_2Pd_3B$ (129.70 states/Ry per unit cell) reported in our previous work.[21] (Same tendency was also reported by S.K. Bose and E.S. Zijlstra.[27]) One can notice that there is a slight quantitative disagreement between the experiment and the calculation on the valence band of $Li_2Pt_3B$, namely the peak position of Pt $d$-bands. Considering the fact that the calculated spectrum of $Li_2Pd_3B$ is in excellent agreement with the experiment,[21] we can possibly consider that spin-orbit coupling plays a major role here. The splitting of bands due to ASOC is expected to cause the increase in bandwidth with negligible shift of spectral center in $Li_2Pt_3B$,[22] and our observation might have captured this effect.

Figure 1 also shows valence band spectra of platinum, $Li_2Pd_{1.5}Pt_{1.5}B$ measured at 13K with 1300eV photon energy together with $Li_2Pt_3B$ and $Li_2Pd_3B$ for comparison. Valence band of platinum is overall consistent with the previous experimental report,[28] but our result shows a clear peak structure near $E_F$ probably due to the better energy resolution and the lower measuring temperature. The observation of high intensity at $E_F$ is expected from the band calculation[28] and the observed asymmetry in Pt $4f$ core levels.[29] Among the samples of $Li_2(Pd_xPt_{1-x})_3B$, intensities at $E_F$ do not show any distinct qualitative difference. Doping of platinum into the system seems to affect the peak positions of feature **B** and **C** as well as the bandwidth, which seems to expand as more platinum is introduced. Moreover, while electron correlation effect is expected to be smaller in $Li_2Pt_3B$, our observation surely shows there is no anomalous feature that invokes strong correlation effect. Therefore, introduction of platinum causes systematic, smooth evolution of electronic structures and the assumption of negligible correlation effect on electronic structure of $Li_2Pt_3B$ is experimentally supported.

Fig. 3 shows a Pd $3d$ core level spectrum of $Li_2Pd_3B$ and $Li_2Pd_{1.5}Pt_{1.5}B$ at 13K with 1300eV photon energy; palladium spectrum from Ref.21 is also shown for comparison. One may notice the symmetry of peak shape in $Li_2Pd_{1.5}Pd_{1.5}B$ and $Li_2Pd_3B$ despite the asymmetry of palladium peaks. The asymmetry of Pd metal peaks should be accounted for the large DOS at $E_F$ [29] as observed in other metallic substances. On the other hand, the symmetric line shape of $Li_2Pd_{1.5}Pt_{1.5}B$ and $Li_2Pd_3B$ indicates the smaller DOS at $E_F$, consistent with our valence band observation. Binding energy of Pd $3d$ state in $Li_2Pd_{1.5}Pt_{1.5}B$ is lower than that of $Li_2Pd_3B$ by 190 meV, and we will discuss this observation later in the text. We could successfully fit the main peaks of Pd $3d$ spectrum of $Li_2Pd_{1.5}Pt_{1.5}B$ with a single asymmetric Gaussian-Lorentzian sum function. This may imply that incorporation of platinum in our sample causes no other significant effect in the chemical bonding states of palladium within the range of our total energy resolution. Moreover, as monovalency is discussed in $Li_2Pd_3B$,[21] peak position of $Li_2Pd_{1.5}Pt_{1.5}B$ close to that of $Li_2Pd_3B$ suggests that effective proximity to $4d^9$ configuration of palladium is possibly maintained in $Li_2Pd_{1.5}Pt_{1.5}B$. We also observed a charge transfer satellite in the $Li_2Pd_{1.5}Pd_{1.5}B$ spectrum, which resembles the one for $Li_2Pd_3B$. It is well known that theoretical analysis of such satellite can give the estimations of charge transfer energy ($\Delta$), hybridization strength ($T$), or on-site Coulomb potential ($U$).[30] The observed similarity in location and intensity of the satellites suggests that $\Delta$, $T$, and $U$ of $Li_2Pd_{1.5}Pd_{1.5}B$ are comparable to these of $Li_2Pd_3B$.

In Fig. 4, we illustrate the Pt $4f$ core level spectra of platinum, $Li_2Pd_{1.5}Pt_{1.5}B$, and $Li_2Pt_3B$ measured at 13K using the photon energy of 1300 eV. We confirm that the platinum spectrum is consistent with the previous XPS studies.[31] The spectra show that Pt $4f$ level of $Li_2Pt_3B$ has a higher binding energy than that of Pt metal, and weak asymmetry of $Li_2Pt_3B$ peak reasonably suggests the decrease of DOS at $E_F$ compared with that of Pt metal,[29] having been confirmed by the valence band measurement. The main peaks of Pt $4f$ spectra were successfully fitted by a single asymmetric Gaussian-Lorentzian sum function, and $Li_2Pd_{1.5}Pt_{1.5}B$ seems to have a slightly higher binding energy of 50 meV. We would like to consider this chemical shift Pt 4*f*, together with Pd 3d observation, in the next paragraph. We also point out that charge transfer satellites were hardly observed in both $Li_2Pd_{1.5}Pt_{1.5}B$ and $Li_2Pt_3B$, but a small satellite was observed in Pt metal spectrum at about 7 eV from the main line. We can utilize these results to estimate the upper limit of *U/W* in $Li_2Pt_3B$, where *U* is an on-site Coulomb potential and *W* is a bandwidth. For this purpose, we tried exact diagonalization calculations for Pt 4*f* spectrum using a single-band Hubbard model with three dimensional cubic lattice structure (4x4x4 sites, 1/16 electrons per site). We find that a small satellite feature arises when *U* is ~3 eV, and since the observed *d*-bandwidth of valence structures in $Li_2Pd_{1.5}Pt_{1.5}B$ and $Li_2Pt_3B$ is about 8 eV (as shown in Fig. 1), we estimate that *U/W* is smaller than 1. To see whether this estimation is reasonable, let's consider previous XPS works on insulating PdO.[32] In T. Uozumi *et al*. (Ref. 32), *U* (on-site Coulomb energy) for PdO is estimated as 4.5 eV. Since *U* in metallic $Li_2Pt_3B$ is expected to be smaller in more localized insulating system of palladium, upper limit of *U* ~ 3 eV in metallic $Li_2Pd_{1.5}Pt_{1.5}B$ and $Li_2Pt_3B$ is reasonable. This can be another experimental confirmation for the absence of strong electron correlations in $Li_2Pd_{1.5}Pt_{1.5}B$, and $Li_2Pt_3B$.

      Figure 5 illustrates B 1*s* spectra of $Li_2Pd_{1.5}Pt_{1.5}B$, and $Li_2Pt_3B$ together with data of $Li_2Pd_3B$. B 1*s* spectra show systematic shifts toward higher binding energy side as Pt content increases (about 100meV in $Li_2Pd_{1.5}Pt_{1.5}B$ and about 200meV in $Li_2Pt_3B$ relative to B 1*s* of $Li_2Pd_3B$). We are to discuss this chemical shift together with the observation in Pd 3*d* and Pt 4*f*. The shift of binding energy, $\Delta E_B$, is given by the following formula:[26, 33] $\Delta E_B = \Delta\mu + K\Delta Q + \Delta V_M - \Delta E_R$. ($\Delta\mu$: the change in chemical potential; *K*: Coulomb coupling constant between the valence and core electrons; $\Delta Q$: the change in the number of valence electrons on the atom of interest; $\Delta V_M$: the change in Madelung potential; and $\Delta E_R$: the change in the extra-atomic relaxation energy) In Li 1*s* (not shown), spectra for $Li_2Pd_{1.5}Pt_{1.5}B$ and $Li_2Pt_3B$ have negligible change within 30 meV, indicating that the change in chemical potentials does not contribute to the shift of B 1*s*. Moreover, the contribution of Madelung potential is unlikely because $Li_2(Pd_xPt_{1-x})_3B$ is a metallic system. Assuming that partial DOS of B 2*s* is negligible at $E_F$ as our calculation indicates, we expect that screening effect in B 1*s* of $Li_2Pt_3B$ is low. Therefore, major contribution to the chemical shift of B 1*s* may be the change in the valency with the incorporation of Pt into the system. Calculated effective charges[22] and the corresponding shifts of Pd 3*d* and Pt 4*f* spectra toward lower binding energy side seems to ensure this interpretation. We can possibly attribute this systematic shift of B 1*s* to the difference in

electronegativity between Pd and Pt, where Pt has the slightly bigger value; however, note that the difference in screening effect is not fully excluded.

In our experimental results, we observed no anomalies that invoke strong electron correlation effect, which is generally unexpected but has to be confirmed experimentally owing to the unconventional superconductivity suggested in $Li_2Pt_3B$. Moreover, our valence band measurement may have directly probed the effect of antisymmetric spin-orbit coupling on the band structure of $Li_2Pt_3B$, but further experimental confirmations are necessary. We could speculate that our results do not conflict with the idea of unconventional superconductivity in $Li_2Pt_3B$ driven by antisymmetric spin-orbit coupling together with the absence of inversion symmetry. We hoped that our first photoemission work on $Li_2Pt_3B$ had motivated more experimental works toward the understanding of superconductivity in $Li_2(Pd_xPt_{1-x})_3B$.

## 4. Conclusion

We have presented the results of the photoemission spectroscopy on the $Li_2Pt_3B$ and $Li_2Pd_{1.5}Pt_{1.5}B$. Our valence band measurement with $Li_2Pt_3B$ shows two major peaks at 2.1 eV and 4.7 eV, the structure at 6.5 eV, and the small non-correlation induced peak at 9.5 eV, and this experimental result was in agreement with the band calculation. By comparison to the calculated partial DOS, we could successfully assign the observed spectral features of $Li_2Pt_3B$ and suggested that the DOS at $E_F$ is predominantly contributed from Pd $5d$ and B $2p$ orbitals. Doping of platinum into the system of $Li_2(Pd_xPt_{1-x})_3B$ simply causes the widening of two major peaks and the expansion of $d$-bandwidth. We observed a slight quantitative discrepancy between the experiment and the calculation in Pt $d$ band of $Li_2Pt_3B$. Our direct observation of valence electronic states may have captured the expected influence of antisymmetric spin-orbit coupling on the band structure of $Li_2Pt_3B$. In Pd $3d$ spectra of $Li_2Pd_{1.5}Pt_{1.5}B$, no variation in chemical environment was suggested, and the observed satellite structure similar to the one in $Li_2Pd_3B$ indicates the set of parameters (charge transfer energy: $\Delta$, hybridization strength: $T$, and on-site Coulomb potential: $U$) defining the electronic state of Pd $d$ bands is comparable among them. Pt $4f$ spectra indicate no large variation of the chemical bonding state in $Li_2(Pd_xPt_{1-x})_3B$. We observed a small satellite structure in Pt metal, but not in $Li_2Pd_{1.5}Pt_{1.5}B$ and $Li_2Pt_3B$. The Hubbard model analysis of Pt $4f$ spectra gives the upper bound estimate of $U\sim3$ eV in the system of $Li_2(Pd_xPt_{1-x})_3B$, giving rise to $U/W<1$. We also presented B $1s$ spectra, and the origin of chemical shift in Pd $3d$, Pt $4f$, and B $1s$ is discussed to be the change in valency.

**Acknowledgements**

We would like to thank K. Iwai and K. Noami for their technical supports. The x-ray photoemission experiments were performed at SPring-8 under the proposal number of J04A25SU-0511N and 2007B1519. This work was partially supported by Grant-in-Aid for Attractive Education in Graduate School "Training Program for Pioneers of Frontier and Fundamental Sciences" from the Ministry of Education, Culture, Sports, Science and Technology.

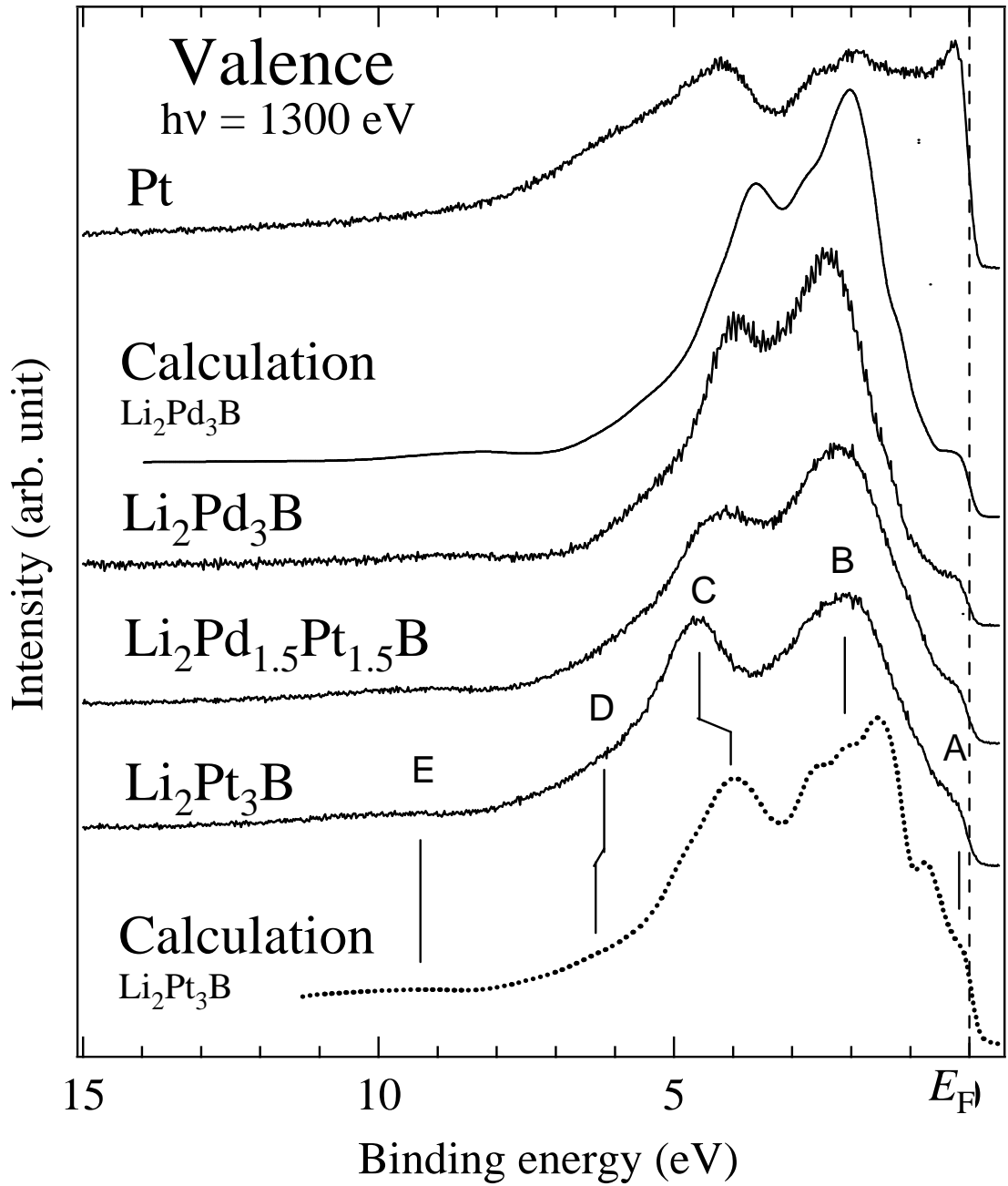

Fig. 1. Valence band spectra of Pt, $Li_2Pd_3B$, $Li_2Pd_{1.5}Pt_{1.5}B$, and $Li_2Pt_3B$ measured at 13 K with 1300 eV photon energy, which are normalized to the area. The band structure calculations of $Li_2Pt_3B$, convoluted with an energy dependent Lorentzian and a Gaussian function, is also shown.

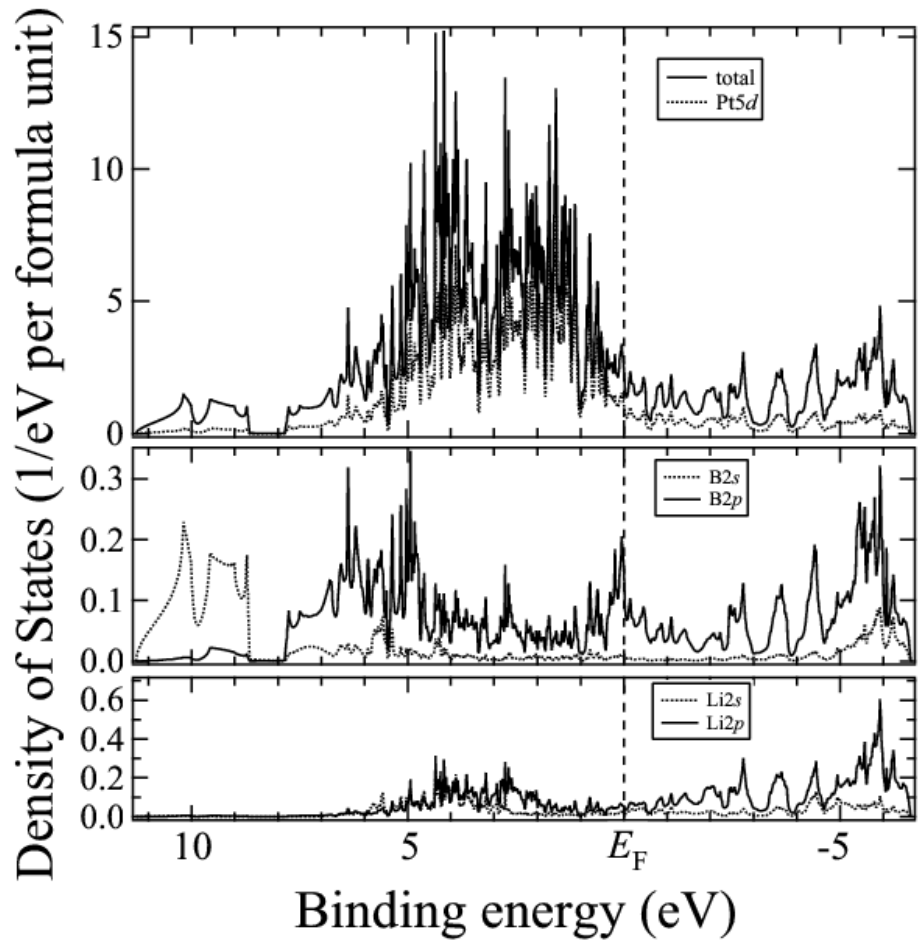

Fig. 2. Calculated total and partial densities of states of $Li_2Pt_3B$.

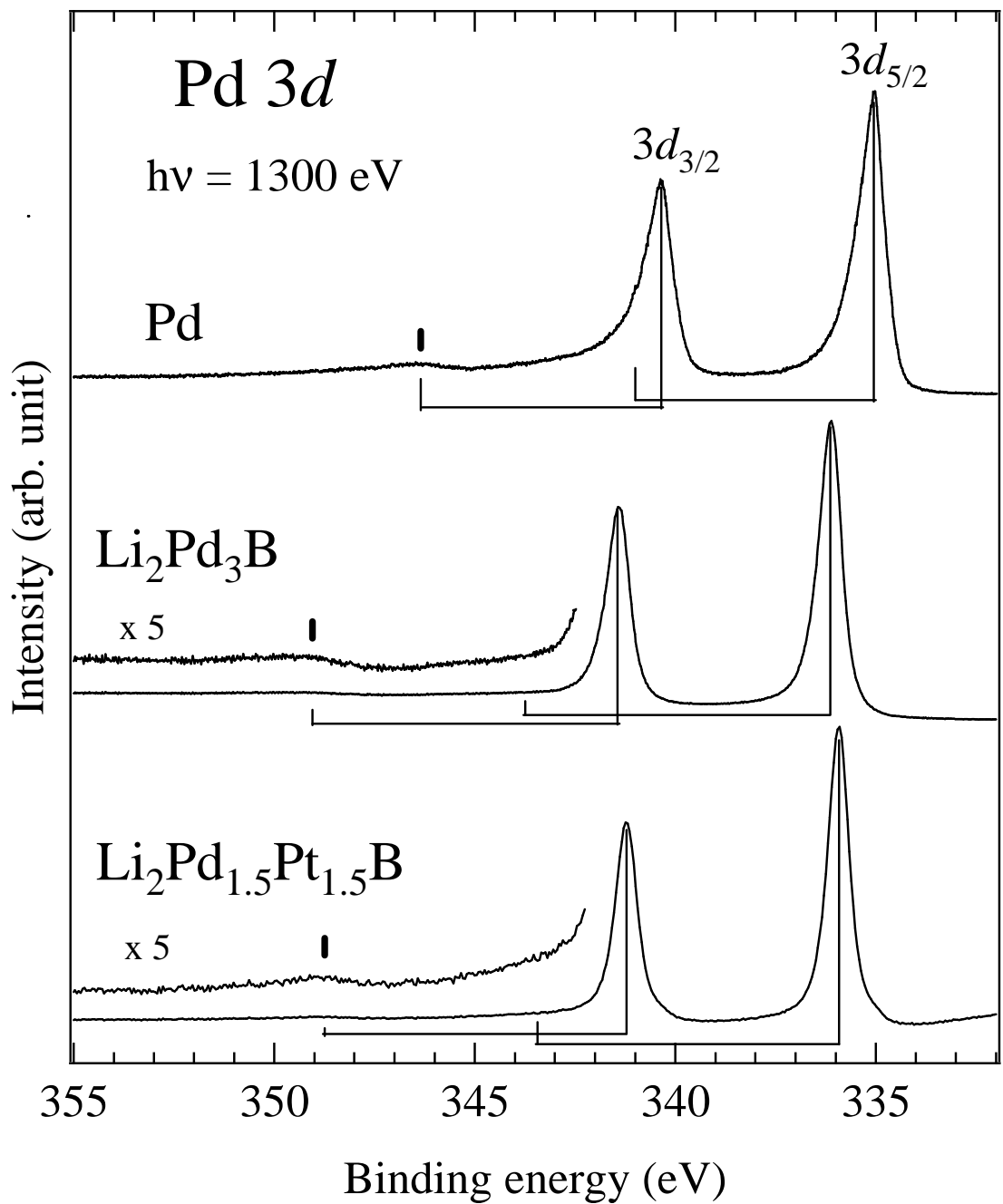

Fig. 3. Pd 3$d$ core level spectrum of Li$_2$Pd$_{1.5}$Pt$_{1.5}$B and Li$_2$Pd$_3$B measured at 13 K with 1300 eV photon energy, together with the data of Pd from Ref. 21. The spectra are normalized to the intensity of 3$d_{5/2}$ peak.

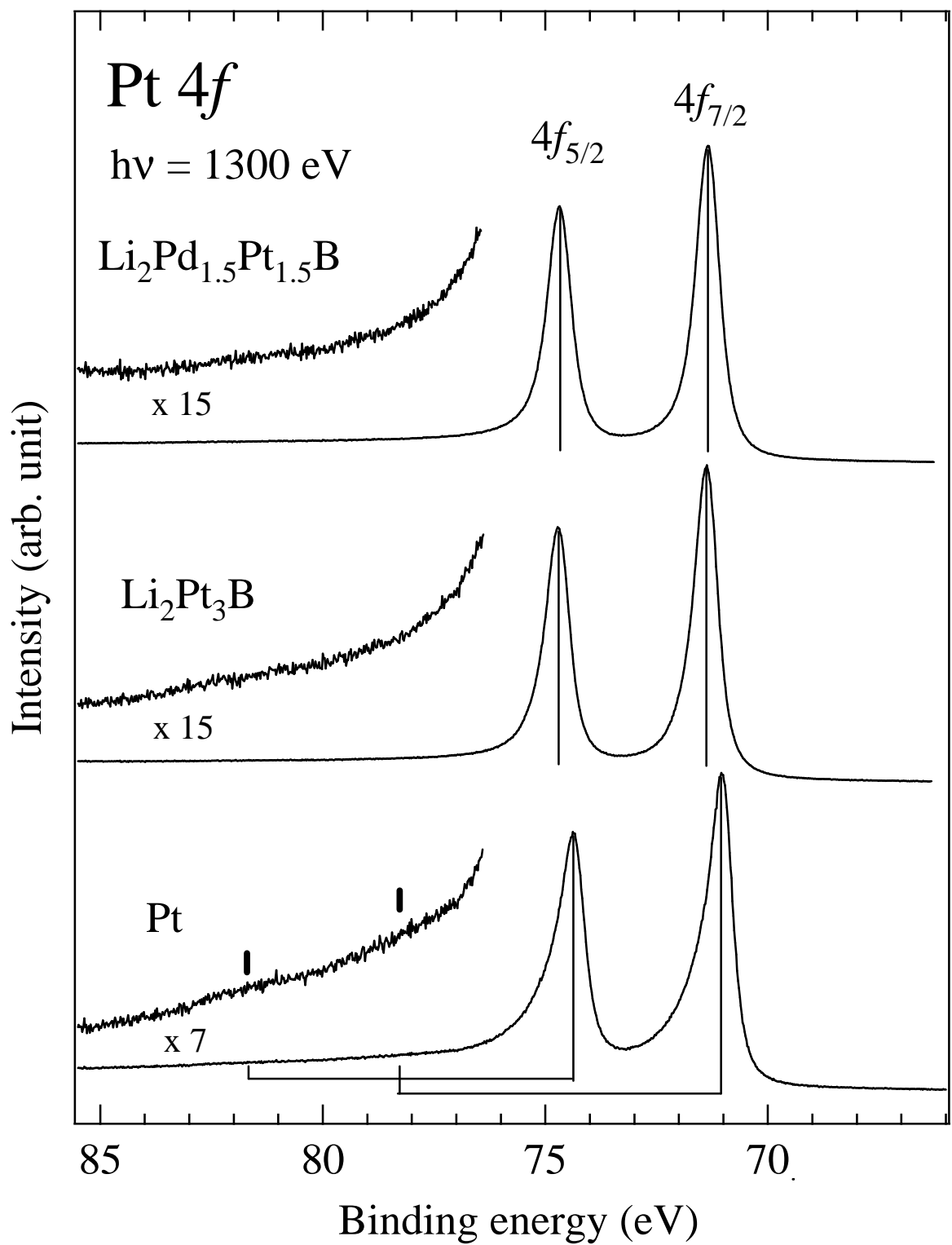

Fig. 4. Pt 4*f* core level spectra of Pt, Li$_2$Pd$_{1.5}$Pt$_{1.5}$B, and Li$_2$Pt$_3$B measured at 13 K with 1300 eV photon energy. The spectra are normalized to the intensity of 4$f_{7/2}$ peak.

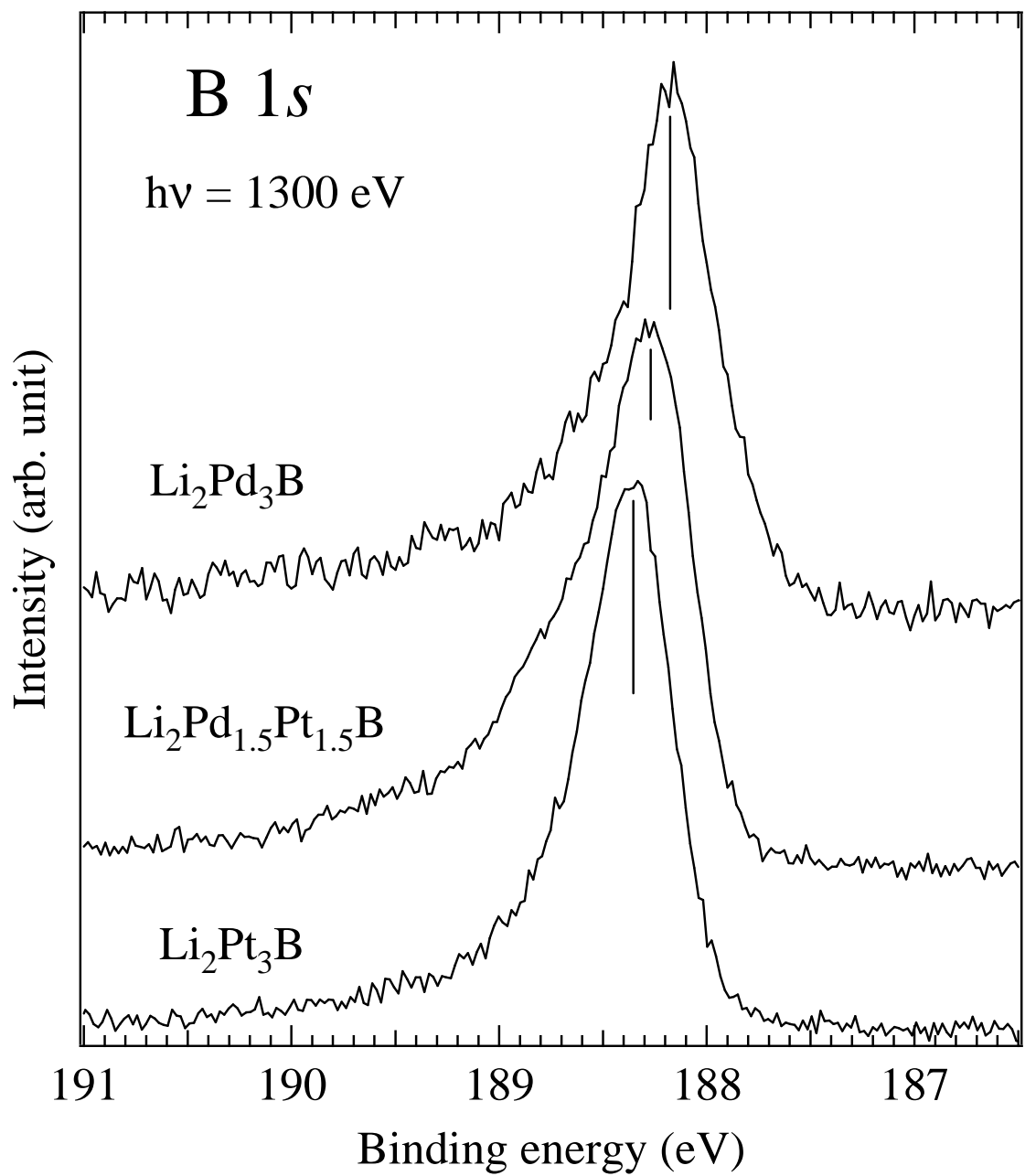

Fig. 5. B 1*s* core level spectra of $Li_2Pd_3B$, $Li_2Pd_{1.5}Pt_{1.5}B$, and $Li_2Pt_3B$ measured at 13 K with 1300 eV photon energy. The spectra are normalized to the intensity of the peak.